\documentclass[pra,aps,preprint,showpacs]{revtex4} 
\newcommand{\ket}[1]{|#1\rangle}      
\newcommand{\bra}[1]{\langle #1|}      
      
\begin{document}       
\title{A note on the geometric phase in adiabatic approximation}       
\author{D. M. Tong$^{a,b,}$\footnote{E-mail address: phytdm@nus.edu.sg}, K. Singh$^a$, L. C. Kwek$^{a,c}$, X. J. Fan$^b$, and C. H. Oh$^{a,}$\footnote{E-mail address: phyohch@nus.edu.sg}}       
\affiliation{$^a$Department of Physics, National University of       
Singapore, 10 Kent Ridge Crescent, Singapore 119260, Singapore \\       
$^b$Department of Physics, Shandong Normal University, Jinan 250014, P. R. China\\
$^c$National Institute of Education, Nanyang Technological       
University, 1 Nanyang Walk, Singapore 639798, Singapore}  
\date{\today}       
\begin{abstract}   
The adiabatic theorem shows that the instantaneous eigenstate is a good approximation of the exact solution for a quantum system in adiabatic evolution. One may therefore expect that the geometric phase calculated by using the eigenstate should be also a good approximation of exact geometric phase. However, we find that the former phase may differ appreciably from the latter if the evolution time is large enough.        
\end{abstract}       
\maketitle       
\date{\today}       
 
The existence of Berry phase was demonstrated in 1983\cite{Berry}. In essence, Berry showed that a quantal system in an instantaneous eigenstate of its Hamiltonian, when slowly transported round a circuit by varying parameters in its Hamiltonian, will acquire a geometric phase factor in addition to the familiar dynamical phase factor. Here he made explicit use of the adiabatic theorem which states that if a quantum system with a time-dependent nondegenerate Hamiltonian $H(t)$ is initially in $n$-th instantaneous eigenstate of $H(0)$, and if $H(t)$ evolves slowly enough, then the state of the system at time $t$ will remain in the $n$-th instantaneous eigenstate of  $H(t)$ up to a multiplicative phase factor\cite{Born,Schwinger,Kato}. This discovery has prompted a myriad of activities that has led to its generalization. Indeed, Anandan and Aharonov\cite{Aharonov} 
extended Berry's result to nonadiabatic cyclic evolutions. J. Samuel and R. Bhandari\cite{Samuel} further generalized the geometric phase to noncyclic and nonunitary evolution. Essentially, the original definition of geometric phase found by Berry for quantum systems under conditions of adiabatic, unitary and cyclic evolutions has been generalized to nonadiabatic, nonunitary and noncyclic evolutions\cite{Mukunda,pati95}. Other generalizations include  geometric phases for mixed states, geometric phase for open systems and off-diagonal geometric phases\cite{Sjoqvistm,Kuldip,manini00,filipp03,carollo03a,Tong,Yi}. 

Recently, some authors\cite{Marzlin,Sarandy,Pati} noted that one must be  careful in employing the adiabatic theorem. Indeed, Marzlin and Sanders demonstrated that a perfunctory application of the theorem may lead to an inconsistency\cite{Marzlin}; and Pati and Rajagopal observed that the Berry phase might be zero if some terms like $\bra {E_m}\dot E_k\rangle$\cite{Pati} are incorrectly dropped and they pointed out that these terms contribute to the Berry phase. These works led us to ask whether there is any significant difference between the approximate phase evaluated under the assumption of adiabacity and the exact geometric phase even if the adiabatic theorem is carefully used. 

In this note, we compare the geometric phase calculated under the assumption of adiabatic theorem with the exact geometric phase. In  examining a well-studied example of a spin-half system in the presence of a rotating magnetic field, we find that the phase calculated by using the adiabatic approximation may differ appreciably from the exact geometric phase if the evolution time is large enough. We further show that the difference is inherently present in general systems. 

\vskip 0.3 cm  

Consider a closed $N$-dimensional quantum system $S$ with the Hamiltonian $H(t)$. The  instantaneous nondegenerate eigenvalues and orthonormal eigenstates of $H(t)$, denoted as $E_m(t)$ and $\ket{E_m(t)}$ respectively, are defined by
\begin{eqnarray}       
H(t)\ket{E_m(t)}=E_m(t)\ket{E_m(t)},~m=1,\ldots,N,
\label{H}       
\end{eqnarray}
Suppose the system is initially in one of the eigenstates $\ket{\psi(0)}=\ket{E_k(0)}$ and its time dependent state is denoted by $\ket{\psi(t)}$. The geometric phase $\gamma(\tau)$ obtained by the system after experiencing an evolution from $t=0$ to $t=\tau$ can be calculated by using the well-known formula\cite{Mukunda}, 
\begin{eqnarray}  
\gamma(\tau)= \arg\langle\psi(0)\ket{\psi(\tau)}+i\int_0^\tau\bra{\psi(t)}\dot {\psi}(t)\rangle dt,
\label{gamma}
\end{eqnarray}  
which is valid for both nonadiabatic as well as adiabatic evolutions.

Specially, if the Hamiltonian $H(t)$ changes slowly and $E_m(t)$ obey the adiabatic constraints
\begin{eqnarray}  
\left |\frac{\langle{E_m(t)}\ket{\dot E_k(t)}}{E_m(t)-E_k(t)}\right | \ll 1,~m\neq k,
\label{constraints}
\end{eqnarray}
according to the adiabatic theorem, the state of the system at time $t$ will be 
\begin{eqnarray}  
\ket{\psi(t)}\approx e^{i\alpha (t)}\ket{E_k(t)},
\label{psit}
\end{eqnarray}
where $\alpha(t)$ is a phase parameter. Then, by substituting Eq. (\ref{psit}) into Eq. (\ref{gamma}), we get the geometric phase in the using of adiabatic theorem as  
\begin{eqnarray}  
\gamma(\tau)= \arg\langle{E_k(0)}\ket{E_k(\tau)}+i\int_0^\tau\bra{E_k(t)}\dot E_k(t)\rangle dt,
\label{gamma2}
\end{eqnarray}
which may be regarded as  the generalized  noncyclic Berry phase, extended from Berry's original definition of cyclic Berry phase\cite{Pati2}.  
It is worth noting that the Berry phase takes the form of
\begin{eqnarray}  
\gamma(\tau)= i\int_0^\tau\bra{E_k(t)}\dot E_k(t)\rangle dt
\label{gamma3}
\end{eqnarray}
if and only if $\arg\langle{E_k(0)}\ket{E_k(\tau)}=0$. 
We would like to stress that Eq. (\ref{gamma2}) is valid for both cyclic and noncyclic adiabatic evolutions, while Eq. (\ref{gamma3}) is valid only when  $\arg\langle{E_k(0)}\ket{E_k(\tau)}=0$.

\vskip 0.3 cm 

It is instructive to ask whether there is any appreciable difference between the approximate  geometric phase and the exact geometric phase for the same system under adiabatic evolution. In other words, we would like to compare the phase calculated using Eq. (\ref{gamma2}), where the adiabatic theorem is applied, with the phase defined by Eq. (\ref{gamma}) with the exact state.
The widely accepted notion is that there is little or no difference between the two. Astonishingly, our result shows that the difference is appreciable if the observing time interval is large enough. 
 
Let us first consider a well-known model; a spin-half particle in a rotating magnetic field, which was also considered  in \cite{Pati}. The Hamiltonian of the model can be  written simply as 
\begin{eqnarray}
H(t)&=&-\frac{\omega_0}{2}(\sigma_x\sin\theta\cos\omega t+\sigma_y\sin\theta\sin\omega t+\sigma_z\cos\theta)\nonumber\\
&=&-\frac{\omega_0}{2}\left(\begin{array}{cc}
\cos\theta&\sin\theta e^{-i\omega t}\\
\sin\theta e^{i\omega t}&-\cos\theta\end{array}\right),
\end{eqnarray} 
where $\omega_0$ is a time-independent parameter defined by the magnetic moment of the spin and the intensity of external magnetic field, $\omega $ is the rotating frequency of the magnetic field and $\sigma_i,~i=x,y,z,$ are Pauli matrices. The instantaneous eigenvalues and eigenstates of $H(t)$ are 
\begin{eqnarray}
E_1=\frac{\omega_0}{2}, ~~E_2=-\frac{\omega_0}{2},
\end{eqnarray} 
and
\begin{eqnarray}  
\ket{E_1(t)}&=&\left(\begin{array}{c}
e^{-i\omega t/2}\sin\frac{\theta}{2}\\-e^{i\omega t/2}\cos\frac{\theta}{2}
\end{array}\right),\nonumber\\
\ket{E_2(t)}&=&\left(\begin{array}{c}
e^{-i\omega t/2}\cos\frac{\theta}{2}\\e^{i\omega t/2}\sin\frac{\theta}{2}
\end{array}\right),
\label{ketE2}
\end{eqnarray}
respectively. The adiabatic constraints (\ref{constraints}) are satisfied as long as
\begin{eqnarray}
 \omega_0\gg \omega\sin\theta.
\label{omegall}
\end{eqnarray}
Suppose that the system is initially in the state $\ket{E_1(0)}$. At time $t$, according to the adiabatic theorem, it will be in the instantaneous eigenstate $\ket{E_1(t)}$ up to a phase factor. The geometric phase for the evolution $\ket{E_1(t)},~t\in[0,\tau],$ can be calculated by Eq. (\ref{gamma2}), and we have
\begin{eqnarray}
\gamma(\tau)=\arg\left(\cos\frac{\omega \tau}{2}+i\cos\theta \sin\frac{\omega \tau}{2}\right)-\frac{1}{2}\omega\tau\cos\theta.
\label{example}
\end{eqnarray}
When $\tau = T = 2\pi/{\omega}$, the cyclic geometric phase reduces to $\gamma(T)=\pi(1-\cos\theta)$.

Next, we evaluate the geometric phase of the above model without any approximation. Here, we solve the Schr\"odinger equation
\begin{eqnarray}  
i\frac{d}{dt}\ket{\psi(t)}=-\frac{\omega_0}{2}\left(\begin{array}{cc}
\cos\theta&\sin\theta e^{-i\omega t}\\
\sin\theta e^{i\omega t}&-\cos\theta\end{array}\right)\ket{\psi(t)},
\label{schrodinger2}
\end{eqnarray} 
with the initial condition $\ket{\psi(0)}=\ket{E_1(0)}$, and obtain the exact solution,
\begin{eqnarray}  
\ket{\psi(t)}=a(t)\ket{E_1(t)}+b(t)\ket{E_2(t)},
\label{solution2}
\end{eqnarray} 
where $\ket{E_1(t)}$, $\ket{E_2(t)}$ are given by Eq. (\ref{ketE2}) and $a(t)$, $b(t)$ are given as
\begin{eqnarray}  
a(t)&=&\cos\frac{\overline{\omega}t}{2}-i\frac{\omega_0+\omega\cos\theta}{\overline{\omega}}\sin\frac{\overline{\omega}t}{2},\nonumber
\\
b(t)&=&i\frac{\omega\sin\theta}{\overline{\omega}}\sin\frac{\overline{\omega}t}{2},
\label{atbt}
\end{eqnarray} 
with $\overline{\omega}=\sqrt{\omega_0^2+\omega^2+2\omega_0\omega\cos\theta}$.
Substituting Eq. (\ref{solution2}) into Eq. (\ref{gamma}), we obtain the geometric phase,
\begin{widetext}
\begin{eqnarray}  
\gamma(\tau)&=&\arg\left[\left(\cos\frac{\overline{\omega}\tau}{2}\cos\frac{\omega\tau}{2}+\frac{\omega+\omega_0\cos\theta}{\overline{\omega}}\sin\frac{\overline{\omega}\tau}{2}\sin\frac{\omega\tau}{2}\right)\right.\nonumber\\&&
+\left.i\left(\cos\theta\cos\frac{\overline{\omega}\tau}{2}\sin\frac{\omega\tau}{2}-\frac{\omega_0+\omega\cos\theta}{\overline{\omega}}\sin\frac{\overline{\omega}\tau}{2}\cos\frac{\omega\tau}{2}\right)\right]\nonumber\\
&&+\frac{\omega_0\omega^2\sin^2\theta}{2\overline{\omega}^3}\sin\overline{\omega}\tau+\frac{\omega_0\tau}{2}-\frac{\omega_0\omega^2\sin^2\theta}{2\overline{\omega}^2}\tau
\label{gammaG}
\end{eqnarray} 
\end{widetext}
Eq. (\ref{gammaG}) gives the exact geometric phase of the system without any approximation. It is valid both for adiabatic evolution as well as for nonadiabatic evolution.  Now, if we invoke the adiabatic condition,  $\omega_0\gg\omega\sin\theta$, we have
\begin{eqnarray}  
\gamma(\tau)&\simeq&\arg\left(\cos\frac{\omega\tau}{2}
+i\cos\theta\sin\frac{\omega\tau}{2}\right)\nonumber\\&-&\frac{1}{2}\omega\tau\cos\theta-\frac{\omega_0\omega^2\sin^2\theta}{2\overline{\omega}^2}\tau.
\label{gammaG2}
\end{eqnarray} 
If $(\omega_0\omega^2\tau\sin^2\theta)/(2\overline{\omega}^2)\ll 1$, we further have
\begin{eqnarray}  
\gamma(\tau)&\simeq&\arg\left(\cos\frac{\omega\tau}{2}
+i\cos\theta\sin\frac{\omega\tau}{2}\right)\nonumber\\&-&\frac{1}{2}\omega\tau\cos\theta,
\label{gammaG3}
\end{eqnarray}
which is the same with Eq. (\ref{example}).
From the above, we find that both cyclic phase $\gamma(2\pi/\omega)$ or noncyclic phase $\gamma(\tau)$ do correctly describe the geometric phase in adiabatic evolution if the time $\tau$ is not too large, say $\tau\in(0,2\pi/\omega)$. We also note that the phase defined by 
Eq. (\ref{gamma2}) may not give the correct geometric phase even in adiabatic evolution if the time $\tau$ is large enough, in which case, the difference between the exact and the approximate value, $\delta\gamma=-(\omega_0\omega^2\tau\sin^2\theta)/(2\overline{\omega}^2)$, can become significant. To elucidate this, we note that with $\overline{\omega}^2\simeq \omega_0^2+2\omega_0\omega\cos\theta$, $\delta\gamma$ can be written as 
\begin{eqnarray} 
\delta\gamma&\simeq& -\tau\cdot\frac{\omega^2\sin^2\theta}{2(\omega_0+2\omega\cos\theta)}\nonumber\\
&=&-\tau s,
\label{small}
\end{eqnarray} 
where  $s=\omega^2\sin^2\theta/2(\omega_0+2\omega\cos\theta)$. Under the adiabatic condition, $s$ is a small quantity. So, $\delta\gamma$ is a small quantity in general. However, $\delta\gamma$ may become an appreciable quantity if $\tau $ is large enough. Although in real experiments $\tau$ is usually taken to be one or two cyclic periods, in theory it may be much larger than one. Note that the adiabatic theorem is valid even for $\tau \gg 2\pi/\omega$ as long as the adiabatic constraints (\ref{omegall}) is satisfied. One may verify that $\ket{\psi(t)}$ in Eq. (\ref{solution2}) always reduces to $\ket{E_1(t)}$ in Eq. (\ref{ketE2}) up to a phase factor, independently of time, as long as the condition (\ref{omegall}) is satisfied. More accurately, one may compare the rays in the projeted Hilbert space. Let $\Delta\rho=\rho_\Psi-\rho_E$, where $\rho_\Psi=\ket{\Psi(t)}\bra{\Psi(t)}$ and $\rho_E=\ket{E_1(t)}\bra{E_1(t)}$. One obtains 
\begin{eqnarray}
\Delta\rho=&&-(\frac{\omega\sin\theta}{\overline\omega})^2\sin^2\frac{\overline\omega t}{2}(\ket{E_1(t)}\bra{E_1(t)}-\ket{E_2(t)}\bra{E_2(t)})\nonumber
\\&&-(\frac{(\omega_0+\omega\cos\theta)\omega\sin\theta}{{\overline\omega}^2}\sin^2\frac{\overline\omega t}{2}+i\frac{\omega\sin\theta}{\overline\omega}\cos\frac{\overline\omega t}{2}\sin\frac{\overline\omega t}{2})\ket{E_1(t)}\bra{E_2(t)}\nonumber\\&&-(\frac{(\omega_0+\omega\cos\theta)\omega\sin\theta}{{\overline\omega}^2}\sin^2\frac{\overline\omega t}{2}-i\frac{\omega\sin\theta}{\overline\omega}\cos\frac{\overline\omega t}{2}\sin\frac{\overline\omega t}{2})\ket{E_2(t)}\bra{E_1(t)}, 
\end{eqnarray}
and will find $\Delta\rho\longrightarrow 0$, in the condition where $\omega_0\gg\omega\sin\theta$.
We may therefore observe that the eigenstate $\ket{E_1(t)}$ is always a good approximation of the exact solution $\ket{\psi(t)}$ for arbitrary times $t$, but the geometric phase calculated by using the eigenstate may not a good approximation of the geometric phase calculated by using the exact solution if $t\gg 2\pi/\omega$.   
 
We now extend the above analysis to a general $N$-dimensional quantum system with Hamiltonian $H(t)$. If the system is initially in the state $\ket{\psi(0)}=\ket{E_k(0)}$ then its exact state at time $t$ can be written as 
\begin{eqnarray}
\ket{\psi(t)}=\sum_mC_m(t)\ket{E_m(t)},
\label{psit3}
\end{eqnarray}
where $C_m(t)$ satisfy $\sum_m C_m^*(t)C_m(t)=1$, $C_m(0)=\delta_{mk}$. Without any loss, we may further let 
\begin{eqnarray}
C_m(t)=\varepsilon_m(t)+\delta_{mk} e^{i\alpha(t)},
\label{cmt}
\end{eqnarray}
where the parameters $\varepsilon_m(t)$ satisfy $\varepsilon_m(0)=0$ and 
\begin{eqnarray}
\sum_m\varepsilon^*_m(t)\varepsilon_m(t)=-\varepsilon_k(t)e^{-i\alpha(t)}-\varepsilon^*_k(t)e^{i\alpha(t)},
\label{normalization}
\end{eqnarray}
and $\alpha(t)$ is defined by Eq. (\ref{psit}). 
Substituting Eqs. (\ref{psit3}) and (\ref{cmt}) into Eq. (\ref{gamma}), we obtain the geometric phase as (see Appendix for details)
\begin{eqnarray}  
\gamma(\tau)&=&\arg\langle{E_k(0)}\ket{E_k(\tau)}+i\int_0^\tau\bra{E_k(t)}\dot E_k(t)\rangle dt\nonumber\\
&&+\Delta \gamma,
\label{gamma5}
\end{eqnarray}
where 
\begin{eqnarray} 
\Delta\gamma&=&\arg\left[1+\sum_m\varepsilon_m(t)\frac{\bra{E_k(0)}E_m(\tau)\rangle}{\bra{E_k(0)}E_k(\tau)\rangle}e^{-i\alpha (\tau)}\right]\nonumber\\&&+i\left(\varepsilon_k(\tau)e^{-i\alpha(\tau)}+\sum_m\int_0^{\varepsilon_m(\tau)}\varepsilon^*_m d\varepsilon_m\right)\nonumber\\
&&-2\text{Re}\left(\int_0^\tau \varepsilon_k(t)\dot \alpha(t)e^{-i\alpha(t)}dt\right)\nonumber\\&&-2\text{Im}\left(\int_0^\tau\sum_m\varepsilon_m(t)\bra{E_k(t)}\dot E_m(t)\rangle e^{-i\alpha(t)}dt\right)\nonumber\\&&-\text{Im}\left(\int_0^\tau\sum_{mn}\varepsilon^*_m(t)\varepsilon_n(t)\bra{E_m(t)}\dot E_n(t)\rangle dt\right).
\label{Dgamma}
\end{eqnarray}  
Eq. (\ref{gamma5}) with Eq. (\ref{Dgamma}) represents the exact geometric phase without any approximation. Notice that the second term on the right side of Eq. (\ref{Dgamma}) is also real, as a consequence of Eq. (\ref{normalization}).

Comparing the phase defined by Eq. (\ref{gamma2}) with the exact geometric phase, we find that $\Delta \gamma $ is just the difference between them. For the quantum system in the adiabatic evolution, $\ket{\psi(t)}\approx e^{i\alpha(t)}\ket{E_k(t)}$, we must have  
$|\varepsilon_m(t)|\ll 1$. $\Delta \gamma$ approximates to zero if the time $\tau$ is not too large, and in this case, the approximate phase does properly describe the geometric phase. However, if the time $\tau$ is large enough, such that $\Delta \gamma$ is not a small quantity, the phase defined by Eq. (\ref{gamma2}) may be very different from the exact geometric phase. This is because the last three terms on the right of Eq. (\ref{Dgamma}) may become appreciably  large as the time progresses unlike the first two terms which can be neglected under the adiabatic approximation. It is important to notice that the property of phase is very different from that of other physical observables. For example, if the number $(2\time10^{10}+1)\pi$ is the value of energy, then it is appropriate to take $2\time10^{10}\pi$ as its approximation; but we cannot do this if it is the value of phase. In other words, an infinitesimal error over a small time interval may accumulate to a measurable quantity as the time becomes large. 

\vskip 0.3 cm  
In conclusion, we have examined the difference between the geometric phase calculated in adiabatic approximation and its exact counterpart. Here we have showed that the difference can become appreciable over long time intervals. This result, in our opinion, is significant, as it debunks the commonly held notion that the two phases are no appreciable difference under the adiabatic approximation. It alerts us that although the adiabatic approximation of wave function is valid, its further application needs to be carful especially in the calculation of geometric phase. This theoretical analysis, however, may not affect existing experimental results since the evolution time in the existing experiments is usually short; within one or two cyclic periods. We also would like to stress that $i\int_0^\tau\bra{E_k(t)}\dot E_k(t)\rangle$ is not generally gauge invariant. Indeed, even in the cyclic adiabatic case, 
the presence of arbitrary phases in the eigenstates of a time-dependent Hamiltonian may introduce non-trivial contribution to the above term.  It should be noted that if the eigenstate satisfies $\arg\langle E_k(0)\ket{E_k(\tau)}=0$, then Eq. (\ref{gamma3}) gives the Berry phase. In general, however, for a system undergoing  adiabatic evolution one should use Eq. (\ref{gamma2}) to calculate the Berry phase.
  
\vskip 0.3 cm     
We thank A. K. Pati for useful remarks. The work was supported by NUS Research Grant No.   
R-144-000-071-305.

\section*{Appendix} 
In the appendix, we outline some of steps involved in  obtaining the expressions (\ref{gamma5}) and (\ref{Dgamma}). 

Substituting Eqs. (\ref{psit3}) and (\ref{cmt}) into $\arg\langle\psi(0)\ket{\psi(\tau)}$, by directly calculating we have
\begin{widetext}
\begin{eqnarray}  
&&\arg\langle\psi(0)\ket{\psi(\tau)}\nonumber\\
&&=\arg\left [\bra{E_k(0)}\sum_{m}C_m(\tau)\ket{E_n(\tau)}\right ]\nonumber\\
&&=\arg\left [\sum_{m}\left(\varepsilon_m(\tau)+\delta_{mk}e^{i\alpha(t)}\right)\bra{E_k(0)}E_m(\tau)\rangle\right ]\nonumber\\
&&=\arg\left [e^{i\alpha(\tau)}\bra{E_k(0)}E_k(\tau)\rangle+\sum_{m}\varepsilon_m(\tau)\bra{E_k(0)}E_m(\tau)\rangle\right ]\nonumber\\
&&=\arg\left [e^{i\alpha(\tau)}\bra{E_k(0)}E_k(\tau)\rangle\left(1+\sum_{m}\varepsilon_m(\tau)\frac{\bra{E_k(0)}E_m(\tau)\rangle}{\bra{E_k(0)}E_k(\tau)\rangle}e^{-i\alpha(\tau)}\right)\right ]\nonumber\\
&&=\alpha(\tau)+\arg\bra{E_k(0)}E_k(\tau)\rangle+\arg\left[1+\sum_{m}\varepsilon_m(\tau)\frac{\bra{E_k(0)}E_m(\tau)\rangle}{\bra{E_k(0)}E_k(\tau)\rangle}e^{-i\alpha(\tau)}\right],
\label{term1}
\end{eqnarray}
where the condition $\bra{E_k(0)}E_k(\tau)\rangle\neq 0$ has been used\cite{Tong3}.  

Substituting Eqs. (\ref{psit3}) and (\ref{cmt}) into $i\int_0^{\tau}\bra{\psi(t)}\dot {\psi}(t)\rangle dt$, we have
\begin{eqnarray}  
&&i\int_0^\tau \bra{\psi(t)}\dot {\psi}(t)\rangle dt\nonumber\\
&&=i\int_0^\tau \sum_{m}C^*_m(t)\bra{E_m(t)}\frac{d}{dt}\left(\sum_nC_n(t)\ket{E_n(t)}\right)dt\nonumber\\
&&=i\int_0^\tau\left[\sum_{m,n}C^*_m(t)\dot C_n(t)\bra{E_m(t)}E_n(t)\rangle+\sum_{m,n}C^*_m(t) C_n(t)\bra{E_m(t)}\dot E_n(t)\rangle\right]dt\nonumber\\
&&=i\int_0^\tau\left[\sum_{m}\left(\varepsilon^*_m(t)+\delta_{mk}e^{-i\alpha(t)}\right)\left(\dot \varepsilon_m(t)+i\delta_{mk}\dot \alpha (t)e^{i\alpha(t)}\right)
\right.\nonumber\\
&&~~\left.+\sum_{m,n}\left(\varepsilon^*_m(t)+\delta_{mk}e^{-i\alpha(t)}\right)\left( \varepsilon_n(t)+\delta_{nk}e^{i\alpha(t)}\right)\bra{E_m(t)}\dot E_n(t)\rangle\right ]dt\nonumber\\
&&=i\int_0^\tau\left[i\dot \alpha(t)+\bra{E_k(t)}\dot E_k(t)\rangle +i\varepsilon^*_k(t)\dot \alpha(t)e^{i\alpha(t)}+\dot \varepsilon_k(t)e^{-i\alpha(t)}+\sum_{m}\varepsilon^*_m(t)\dot \varepsilon_m(t)\right.\nonumber\\
&&~~~+\sum_{m}\varepsilon^*_m(t)e^{i\alpha(t)}\bra{E_m(t)}\dot E_k(t)\rangle+\sum_{n}\varepsilon_n(t)e^{-i\alpha(t)}\bra{E_k(t)}\dot E_n(t)\rangle\nonumber\\
&&~~~\left. +\sum_{m,n}\varepsilon^*_m(t) \varepsilon_n(t)\bra{E_m(t)}\dot E_n(t)\rangle\right]dt\nonumber\\
\end{eqnarray} 
Noting that 
\begin{eqnarray} 
&&i\int_0^\tau\sum_m\varepsilon^*_m(t)\dot \epsilon_m(t) dt=i\sum_m\int_0^{\varepsilon_m(\tau)}\varepsilon^*_m d\epsilon_m ,\nonumber\\
&&i\int_0^\tau\left[i\varepsilon^*_k(t)\dot \alpha(t)e^{i\alpha(t)}+\dot \varepsilon_k(t)e^{-i\alpha(t)}\right]dt =i\varepsilon_k(\tau)e^{-i\alpha(\tau)}-2\text{Re}\left(\int_0^\tau \varepsilon_k(t)\dot \alpha(t)e^{-i\alpha(t)}dt\right) ,\nonumber\\
&&i\int_0^\tau \sum_{m,n}\varepsilon^*_m(t) \varepsilon_n(t)\bra{E_m(t)}\dot E_n(t)\rangle dt=
-\text{Im}\left(\int_0^\tau\sum_{mn}\varepsilon^*_m(t)\varepsilon_n(t)\bra{E_m(t)}\dot E_n(t)\rangle dt\right),\nonumber\\
\text{and}\nonumber\\
&&i\int_0^\tau \left(\sum_{m}\varepsilon^*_m(t)e^{i\alpha(t)}\bra{E_m(t)}\dot E_k(t)\rangle+\sum_n\varepsilon_n(t)e^{-i\alpha(t)}\bra{E_k(t)}\dot E_n(t)\rangle\right)dt\nonumber\\
&&=-2\text{Im}\left(\int_0^\tau\sum_m\varepsilon_m(t)\bra{E_k(t)}\dot E_m(t)\rangle e^{-i\alpha(t)}dt\right),
\end{eqnarray}
we obtain
\begin{eqnarray}
&&i\int_0^\tau \bra{\psi(t)}\dot {\psi}(t)\rangle dt\nonumber\\
&&=-\alpha(\tau)+i\int_0^\tau\bra{E_k(t)}\dot E_k(t)\rangle dt+i\left(\varepsilon_k(\tau)e^{-i\alpha(\tau)}+\sum_m\int_0^{\varepsilon_m(\tau)}\varepsilon^*_m d\varepsilon_m\right)\nonumber\\
&&~~~-2\text{Re}\left(\int_0^\tau \varepsilon_k(t)\dot \alpha(t)e^{-i\alpha(t)}dt\right)
-2\text{Im}\left(\int_0^\tau\sum_m\varepsilon_m(t)\bra{E_k(t)}\dot E_m(t)\rangle e^{-i\alpha(t)}dt\right)
\nonumber\\
&&~~~-\text{Im}\left(\int_0^\tau\sum_{mn}\varepsilon^*_m(t)\varepsilon_n(t)\bra{E_m(t)}\dot E_n(t)\rangle dt\right).
\label{term2}
\end{eqnarray}  
\end{widetext}
Taking Eqs. (\ref{term1}) and (\ref{term2}) into (\ref{gamma}), the expressions (\ref{gamma5}) and (\ref{Dgamma}) are immediately obtained.


\begin{thebibliography}{99}    
\bibitem{Berry} M.V. Berry,    
Proc. R. Soc. London Ser. A {\bf 392}, 45 (1984).    
\bibitem{Born}M. Born and V. Fock,    
Z. Phys. {\bf 51}, 165(1928).
\bibitem{Schwinger}J. Schwinger, Phys. Rev. {\bf 51}, 648(1937).
\bibitem{Kato}T. Kato, J. Phys. Soc. Jap. 5, 435 (1950).
\bibitem{Aharonov} Y. Aharonov and J. Anandan,    
Phys. Rev. Lett. {\bf 58}, 1593 (1987); 
J. Anandan and Y. Aharonov,    
Phys. Rev. D {\bf 38}, 1863 (1988).    
\bibitem{Samuel} J. Samuel and R. Bhandari,    
Phys. Rev. Lett. {\bf 60}, 2339 (1988).    
\bibitem{Mukunda} N. Mukunda and R. Simon,    
Ann. Phys. (N.Y.) {\bf 228}, 205 (1993).    
\bibitem{pati95} A.K. Pati,    
Phys. Rev. A {\bf 52}, 2576 (1995); 
A.K. Pati,    
J. Phys. A {\bf 28}, 2087 (1995).    
\bibitem{Sjoqvistm} E. Sj\"oqvist, A.K. Pati, A. Ekert,   
J.S. Anandan, M. Ericsson, D.K.L. Oi, and V. Vedral,   
Phys. Rev. Lett. {\bf 85}, 2845 (2000).    
\bibitem{Kuldip} K. Singh, D.M. Tong, K. Basu, J.L. Chen, and J.F. Du,   
Phys. Rev. A  {\bf 67}, 032106 (2003).    
\bibitem{manini00} N. Manini and F. Pistolesi,  
Phys. Rev. Lett. {\bf 85}, 3067 (2000).  
\bibitem{filipp03} S. Filipp and E. Sj\"oqvist,  
Phys. Rev. Lett. {\bf 90}, 050403 (2003).  
\bibitem{carollo03a} A. Carollo, I. Fuentes-Guridi, M. F. Santos and  
V. Vedral,  
Phys. Rev. Lett. {\bf 90}, 160402 (2003).
\bibitem{Tong} D.M. Tong, E. Sj\"{o}qvist, L.C. Kwek, and C.H. Oh, Phys. Rev. Lett. {\bf 93}, 080405(2004).
\bibitem{Yi}X. X. Yi, L. C. Wang, and T. Y. Zheng, Phys. Rev. Lett. {\bf 92}, 150406(2004).
\bibitem{Marzlin}Karl-Peter Marzlin and Barry C. Sanders, Phys. Rev. Lett. {\bf 93}, 160408(2004)
\bibitem{Sarandy}M.S. Sarandy, L. A. Wu, and D. A. Lidar, quant-ph/0405059(2004).
\bibitem{Pati}A. K. Pati and A. K. Rajagopal,    
quant-ph/0405129v1(2004).
\bibitem{Pati2}A. K. Pati, Ann. Phys.  {\bf 270}, 178 (1998).   
\bibitem{Tong3} For the standard (diagonal) geometric phase, $\bra{E_k(0)}E_k(\tau)\rangle$ cannot be zero. Otherwise, the phase is undetermined, and the off-diagonal geometric phase needs to be considered.  
\end{thebibliography}
\end{document}